\documentclass{iaus}
\usepackage{graphicx}
\def\apj{{ApJ}}			
\def\apjl{{ApJ}}		
\def\apjs{{ApJS}}

\def\aap{{A\&A}}

\def\pasj{{PASJ}}

\def\solphys{{Solar\ Phys.}}

\def\la{\mathrel{\hbox{\rlap{\hbox{\lower4pt\hbox{$\sim$}}}\hbox{$<$}}}}
\def\ga{\mathrel{\hbox{\rlap{\hbox{\lower4pt\hbox{$\sim$}}}\hbox{$>$}}}}

\def\rmit#1{{\it #1}}              %% italics (RR style, Kluwer)
\def\rmit#1{{\rm #1}}              %% redefine for ASP, A&A, ApJ, Springer??
\def\ie{\rmit{i.e.,}}              %% , required (Webster 1681)
\def\lesssim{\mathrel{\hbox{\rlap{\hbox{\lower4pt\hbox{$\sim$}}}\hbox{$<$}}}}
\def\gtrsim{\mathrel{\hbox{\rlap{\hbox{\lower4pt\hbox{$\sim$}}}\hbox{$>$}}}}
\def\level #1 #2#3#4{$#1 \: ^{#2} \mbox{#3} ^{#4}$}

\usepackage{natbib}

\title[Are the sunspots force-free?] %% give here short title %%
{Are the photospheric sunspots magnetically force-free in nature?}

\author[Sanjiv Kumar Tiwari]   %% give here short author list %%
{Sanjiv Kumar Tiwari}%$^1$}
\affiliation{Udaipur Solar Observatory, Physical Research Laboratory,
Dewali, Bari Road,\\
Udaipur - 313 001, India.
\\ email: {\tt stiwari@prl.res.in}} %\\[\affilskip]
\pubyear{2011}
\volume{273}  %% insert here IAU Symposium No.
\pagerange{1--8}
\jname{IAU-273, Physics of Sun and Star Spots}
\editors{D. P. Choudhary \& K. G. Strassmeier, eds.}
\begin{document}

\maketitle

\begin{abstract}
In a force-free magnetic field, there is no interaction of field and
the plasma in the surrounding atmosphere i.e., electric currents are
aligned with the magnetic field, giving rise to zero Lorentz force.
The computation of many magnetic parameters like magnetic energy,
gradient of twist of sunspot magnetic fields (computed from the
force-free parameter $\alpha$), including any kind of extrapolations
heavily hinge on the force-free approximation of the photospheric
magnetic fields. The force-free magnetic behaviour of the
photospheric sunspot fields has been examined by \cite{metc95} and
\cite{moon02} ending with inconsistent results. \cite{metc95}
concluded that the photospheric magnetic fields are far from the
force-free nature whereas \cite{moon02} found the that the
photospheric magnetic fields are not so far from the force-free
nature as conventionally regarded. The accurate photospheric vector
field measurements with high resolution are needed to examine the
force-free nature of sunspots. We use high resolution vector
magnetograms obtained from the Solar Optical
Telescope/Spectro-Polarimeter (SOT/SP) aboard Hinode to inspect the
force-free behaviour of the photospheric sunspot magnetic fields.
Both the necessary and sufficient conditions for force-freeness are
examined by checking global as well as as local nature of sunspot
magnetic fields. We find that the sunspot magnetic fields are very
close to the force-free approximation, although they are not
completely force-free on the photosphere.
\keywords{Sun: atmosphere, Sun: force-free fields, Sun: magnetic
fields, Sun: sunspots}
\end{abstract}

\firstsection % if your document starts with a section,
\section{Introduction}
A force-free magnetic field does physically mean a zero Lorentz
force \citep{chandra61,parker79,low82} \ie\ ($\nabla\times \bf B)
\times \bf B = 0$. This equation can be rewritten as
\begin{equation}
\nabla\times \bf B = \alpha \bf B
\end{equation}
The z component of above condition allows us to compute the
distribution of $\alpha$ on the photosphere (z=0)
\begin{equation}
\alpha = \left[{\frac{\partial B_y}{\partial x} - \frac{\partial B_x}{\partial y}}\right]/{B_z}
\end{equation}
Three cases may arise: (i) $\alpha$ = 0 everywhere, i.e., no
electric current in the atmosphere resulting in a potential field
(\cite{schm64,semel67,saku89,regn07} (ii) $\alpha$ = constant
everywhere, \ie\ linear force-free state
\cite[etc]{naka72,gary89,van10} which is not always valid and (iii)
$\alpha$ varies spatially, \ie\ nonlinear force-free magnetic field
\citep{saku79,low82a,amar06,wieg04,schr08,derosa09,mack09}, this is
the most common state expected \citep{low85}. However, high
resolution vector magnetograms are required to confirm this. In
earlier works, perhaps the poor resolution of data obscured the
conclusions about the validity of linear/non-linear force-free
approximations. In the present work, we check the validity of
linear/nonlinear assumptions along with examining the force-freeness
over sunspot magnetic fields using high spatial resolution
photospheric vector magnetograms obtained from Solar Optical
Telescope/Spectro-Polarimeter onboard Hinode. The effect of
polarimetric noise present in the data obtained from SOT/SP does not
affect much in derivation of the magnetic field parameters
\citep{tiw09a,gosain10}. \vspace{-.4cm}
\section{Necessary and sufficient conditions}
{\underline{\it Necessary condition}}. Under the assumption that the
magnetic field above the plane z=0 (photosphere) falls off enough as
z goes to infinity, the net Lorentz force in the volume z$>$0 is
just the Maxwell stress integrated over the plane z=0 (Aly, 1984;
Low, 1985). Thus the components of the net Lorentz force at the
plane z=0 can be expressed by the surface integrals as follows:
\begin{equation}\label{loren}
F_x = -\frac{1}{4\pi}\int B_x B_y dxdy; F_y = -\frac{1}{4\pi}\int B_y B_z dxdy; F_z = -\frac{1}{8\pi}\int B^2_z-B^2_x-B^2_y dxdy
\end{equation}
where $F_x$, $F_y$ and $F_z$ represent the components of the net
Lorentz force. According to \cite{low85} the necessary conditions
for any magnetic field to be force-free are that
\begin{equation}\label{cond}
|F_x| << |F_p|; ~~~ |F_y| << |F_p|;~~~ |F_z| << |F_p|
\end{equation}
where $F_p$ is force due to the distribution of magnetic pressure on
z=0, as given by,
\begin{equation}\label{press}
F_p = -\frac{1}{8\pi}\int B^2_z+B^2_x+B^2_y dxdy
\end{equation}
It was discussed by \cite{metc95} that the magnetic field is
force-free if the aforementioned ratios are less or equal to 0.1. It
is to be noted that the conditions \ref{cond} are only necessary
conditions for the fields to be force-free. The reason for this is
that some information is lost in the surface integration in
Equations \ref{loren}.

{\underline{\it Sufficient condition}}. In a force-free case the
tension force will balance the gradient of magnetic pressure
demanding for zero Lorentz force. We can split up the Lorentz force
($F = (1/c)J \times B$) in two terms as,
\begin{equation}\label{lorz}
    {\bf F} = \frac{\bf (B \cdot \nabla) B}{4\pi} - \frac{\bf \nabla (B \cdot B)}{8\pi}
\end{equation}
The first term in the right hand side in the above equation is the
tension force ({\bf T}). The second term represents the gradient of
the magnetic pressure i.e., the force due to magnetic pressure ($\bf
F_p$). The vertical component of the tension force term can be
simplified to,
\begin{equation}\label{tz}
{T_z} = \frac{1}{4\pi}[B_x \frac{\partial B_z}{\partial x} + {B_y} \frac{\partial B_z}{\partial y} -
B_z (\frac{\partial B_x}{\partial x} + \frac{\partial B_y}{\partial y})]
\end{equation}
where, the last component has been drawn from the condition ${\bf
\nabla \cdot B } = 0$. The usefulness of the tension force has not
found much attention earlier in the literature but for few studies
\citep{venk90,venk93,venk10}. Recently \cite{venk10} pointed out the
utility of tension force as a diagnostic of dynamical equilibrium of
sunspots. It was found \citep{venk10} that the magnitude of vertical
tension force attains values comparable to the force of gravity at
several places over the sunspots meaning that the non-magnetic
forces will not be able to balance this tension force. Only gradient
of the magnetic pressure can match this force resulting into the
force-free configurations. This serves as a sufficient condition for
verifying the force-freeness of the sunspot magnetic fields. For
details, see Tiwari, 2010, in prep. \vspace{-.3cm}
\section{Data and analysis}
We have used the high resolution vector magnetograms obtained from
the Solar Optical Telescope/Spectro-polarimeter (SOT/SP:
\cite{tsun08,suem08,ichi08,shim08}) onboard Hinode (\cite{kosu07}).
The data has been prepared as done successfully in
\cite{tiw09b,tiw10a,venk09,venk10,gosain09,gosain10}).
\begin{figure}[h]
 %\vspace*{-2.0 cm}
\begin{center}
 \includegraphics[width=4.2in]{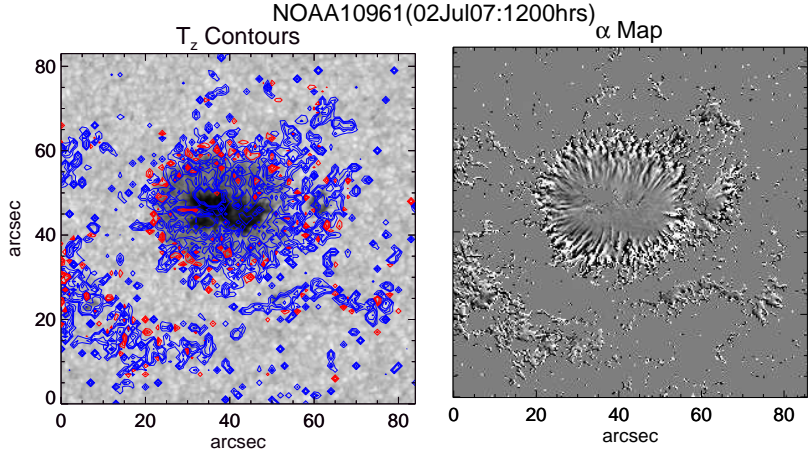}
 %\vspace*{-0.5 cm}
\caption{Left panel: Vertical tension force distribution of
a sunspot NOAA AR 10961 over its continuum map. Blue (red) colors show negative (positive)
contours of $\pm0.4, \pm1.2, \pm4, \pm12$ millidynes/cm$^3$. Right panel: Alpha map of
the same active region.}
\label{fig1}
\end{center}
\end{figure}
\vspace*{-0.3 cm}
\begin{figure}[h]
\begin{center}
\includegraphics[width=2.4in]{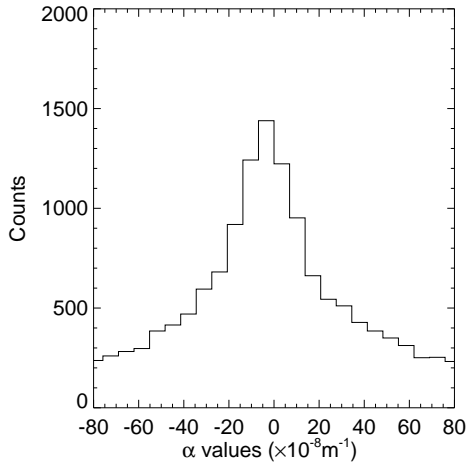}\hspace{2.pc}
\begin{minipage}[b]{7pc}
\vspace{-2.8cm}
\caption{Histogram of the $\alpha$ values of NOAA AR 10961 observed
on July 12, 2007 at 1200UT, as an example. We can see even in this simple
sunspot, the alpha has a wide range of its distribution.}
\end{minipage}
\end{center}
\end{figure}

\vspace{-.4cm}
\section{Results and Discussion}
We conclude the following: 1. the sunspot magnetic fields are not
far from the force-free nature as has been suspected for long time.
2. The non-linear force-free approximation is closer to validity in
all sunspots, either it is a simple active region or a complex one.

It is well known that all extrapolation techniques rely on the
photospheric vector field measurements and also on its force-free
approximation. Coronal magnetic field reconstruction by
extrapolations of photospheric magnetic fields under non-linear
modeling have shown satisfactory results by roughly matching with
the coronal observations \cite{mccl94,wieg05,schr06}. These results
then also support our conclusion that the sunspot magnetic fields
are close to non-linear force-free approximation. Greater details
will be given in Tiwari, 2010a (in preparation). \vspace{-.2cm}
\acknowledgments The presentation of this paper in the IAU Symposium
273 was possible due to  partial support from the National Science
Foundation grant numbers ATM 0548260, AST 0968672 and NASA - Living
With a Star grant number 09-LWSTRT09-0039. Hinode is a Japanese
mission developed and launched by ISAS/JAXA, with NAOJ as domestic
partner and NASA and STFC (UK) as international partners. It is
operated by these agencies in co-operation with ESA and NSC
(Norway).

%\bibliographystyle{apj}
%\bibliography{stiwari}

\end{document}